# Physics-Aware Zero-Shot Learning for Low-Dose CT from a Single Sinogram

Yongyi Shi, Siming Lu, Yongfeng Gao, Hao Zhang, Hengyong Yu, Ge Wang

***Abstract*—Low-dose computed tomography (CT) imaging remains a persistent challenge, especially under heterogeneous acquisition conditions. Zero-shot learning, which enables models to learn directly from individual images without external supervision, represents a compelling approach to mitigate distribution mismatches arising from variations in scanner hardware, imaging protocols, and patient demographics. However, most existing zero-shot approaches employ down-sampling operations that reduce spatial resolution and cause domain shifts, which become particularly evident when applied to raw sinogram data. Here, we propose a physics-aware zero-shot CT denoising framework to exploit the conjugate symmetry property intrinsic to CT sinograms. Conjugate X-ray pencil beams traverse identical anatomical paths and ideally yield the same expected attenuation values but with independent noise realizations. By randomly mapping sinogram values between conjugate-ray pairs, we synthesize structurally consistent yet noise-diverse sinogram pairs, enabling self-supervised network training without external data. The resulting lightweight denoising network demonstrates strong generalization across low-dose simulation, clinical helical CT, and preclinical photon-counting CT datasets. Quantitative and qualitative results demonstrate competitive denoising performance, with favorable noise suppression and structural preservation across diverse acquisition settings. This work establishes a generalizable and data-efficient paradigm for CT denoising that integrates physical redundancy into learning-based image reconstruction. The code for the proposed method is openly available at: https://github.com/shiyongyi/Physics-Aware-Zero-Shot-CT-Sinogram-Denoising.***

## I. Introduction

X-RAY computed tomography (CT) is a cornerstone of modern diagnostic imaging, underpinning disease detection, treatment planning, and image-guided interventions. The ionizing radiation intrinsic to CT has long raised concerns regarding cumulative exposure and elevated cancer risk, driving continued efforts to reduce radiation dose through low-dose CT (LDCT) protocols [1, 2]. Yet reducing dose inevitably increases image noise and degrades image quality, compromising diagnostic reliability [3]. Accordingly, LDCT reconstruction and denoising have emerged as pivotal challenges in medical imaging research [4-7].

Deep learning has revolutionized LDCT denoising in recent years [8-17], with supervised models achieving outstanding performance in suppressing noise and preserving anatomical details. Despite these advances, such models rely on large paired datasets of noisy and clean images that are nearly impossible to obtain in clinical settings, because this would require multiple scans of the same patient at different dose levels. Ethical restrictions, patient motion, and physiological variability further complicate the collection of well-aligned pairs. To address this limitation, many studies generate training data synthetically by adding Poisson or Gaussian noise to normal-dose sinograms [18]. This strategy simplifies data acquisition but introduces distributional discrepancies across scanner types, imaging protocols, and patient populations [19, 20], which in turn weaken model robustness, fairness, and clinical generalizability.

Self-supervised frameworks such as Noise2Noise [21, 22] eliminate the need for clean reference data by training networks on pairs of independently noisy images that share the same underlying signal. When the noise is zero-mean and statistically independent, minimizing the reconstruction loss enables the network to approximate the expected clean image. This assumption holds well in natural image data, where multiple noisy measurements of the same scene can be readily obtained, and has been shown to yield impressive denoising performance. Extending this strategy to CT, however, is far more challenging. Acquiring repeated scans of the same anatomy under independent noise realizations would violate radiation-safety regulations. Some studies have instead attempted to construct pseudo-pairs by subsampling projection views from a single acquisition [23–25]. But this approach alters the noise statistics and breaks the geometric consistency of the projections, leading to structural bias and degraded reconstruction fidelity. As a result, Noise2Noise-style training remains difficult to apply directly to CT data without introducing additional artifacts.

Recent advances have focused on zero-shot denoising, which constructs training pairs directly from a single noisy image, enabling network optimization without any external datasets.

This work was supported in part by the National Institute of Biomedical Imaging and Bioengineering/National Institutes of Health under Grant R01 EB032716. The authors declare that they have no conflicts of interest related to this work. (Corresponding authors: Yongyi Shi, Ge Wang; e-mail: shiyongyi@csu.edu.cn, wangg6@rpi.edu).

Yongyi Shi and Ge Wang are with the Biomedical Imaging Center, Center for Computational Innovations, Center for Biotechnology & Interdisciplinary Studies, Department of Biomedical Engineering, School of Engineering, Rensselaer Polytechnic Institute, Troy, NY 12180 USA.

Hengyong Yu is with the Department of Electrical and Computer Engineering, University of Massachusetts at Lowell, Lowell, MA 01886 USA.

Siming lu and Yongfeng Gao are with the Department of Radiology, Stony Brook University, Stony Brook, NY 11794 USA.

Hao Zhang is with the Department of Radiation Oncology, Stanford University, Stanford, CA 94305 USA.

Noise2Void [26–29] employs a blind-spot network that predicts masked pixel intensities from their spatial neighborhood, assuming noise independence across pixels. Neighbor2Neighbor [30,31] forms pseudo-pairs by splitting neighboring pixels into interleaved subsets, allowing the model to learn local signal consistency under independent noise. Noise2Fast [32] accelerates this process through a checkerboard downsampling scheme that generates four quasi-independent training images from a single input, offering efficient self-supervised denoising. ZS-Noise2Noise [33] further introduces a consistency loss across multiple augmented versions of the same noisy image, improving training stability and robustness under diverse noise conditions.

Despite these advances, the finite training data extracted from a single image and potential resolution degradation due to downsampling pose inherent challenges to achieve optimal generalization and image fidelity, particularly in the sinogram domain of CT. Unlike image-domain representations, sinograms directly encode the original projection data and thus provide richer information for reconstruction, often yielding superior performance over single image-domain methods [34,35]. Furthermore, improving sinogram quality has the potential to improve downstream reconstruction, as higher-quality sinogram could enable more effective use of image-domain deep priors for anatomical detail preservation and noise suppression [36, 37]. These considerations highlight the pivotal role of sinogram denoising to advance low-dose CT imaging [38, 39]. Nevertheless, the acquisition of sinograms is difficult and often non-standardized, making zero-shot approaches especially appealing for sinogram denoising. Meanwhile, current zero-shot methods have not been explored in this domain, largely because sinogram signals are highly sensitive: even small errors in the sinogram can lead to severe artifacts in the reconstructed images [40]. Consequently, it remains highly challenging to transfer zero-shot strategies to the sinogram domain.

To address these challenges, we propose a physics-aware zero-shot learning (PZL) LDCT denoising framework that exploits the inherent conjugate symmetry of CT sinograms, drawing inspiration from the redundancy present in the projection domain [41]. In principle, two conjugate X-ray pencil beams traverse the same anatomical path and are expected to yield identical signal values, although their noise characteristics may differ due to stochastic measurement variations. By randomly mapping portions of these conjugate rays within the sinogram domain, we can synthesize a large set of structurally consistent yet noise-varied sinograms. This conjugate-based augmentation effectively reduces spatial noise correlation and greatly expands the diversity of training data without sacrificing resolution or anatomical information. By operating directly on the entire sinogram rather than relying on restrictive local sampling or masking strategies, our method preserves full anatomical contents, ensures high-fidelity reconstruction, and improves robustness against spatially structured noise, making it highly compatible with clinical LDCT imaging.

Beyond conventional LDCT, we extend our framework to photon-counting CT, a rapidly emerging modality that has transitioned from research prototypes to clinical applications. Photon-counting CT offers distinct advantages such as energy-resolved imaging, improved material differentiation, and reduced electronic noise. Nevertheless, it also presents significant challenges stemming from low photon statistics, charge sharing, and energy-dependent crosstalk, leading to complex and spatially varying noise distributions. Meanwhile, the scarcity of large-scale photon-counting CT datasets imposes a major limitation on supervised learning strategies, whereas most existing reconstruction methods are formulated within iterative optimization frameworks that rely on modality-specific priors or learned regularizers to stabilize spectral and photon-counting reconstruction [42–44], without direct modeling or optimization in the sinogram domain. In this context, it offers unique advantages for photon-counting CT to perform denoising in the sinogram space, as it suppresses measurement noise before reconstruction and enhances the stability of subsequent image formation. Hence, it becomes particularly appealing for a zero-shot strategy that is capable of adapting directly to individual photon-counting CT scans without external training data. The proposed conjugate ray–based augmentation exploits the intrinsic spatial redundancy of photon-counting sinograms to achieve robust and reference-free denoising, representing a promising direction toward data-efficient and generalizable photon-counting CT imaging.

We validate the effectiveness of our framework through extensive experiments on both simulated and real-world datasets, including low-dose CT sinogram denoising and photon-counting CT sinogram denoising. Quantitative and qualitative evaluations demonstrate that our method achieves superior resolution preservation and noise suppression compared to the existing zero-shot denoising techniques, offering a robust and practical solution for clinical imaging applications across multiple modalities.

In summary, this work makes the following contributions:

1) We introduce a zero-shot LDCT denoising framework by leveraging conjugate symmetry to generate structurally consistent yet noise-varied sinogram pairs from a single sinogram.

2) We propose a conjugate ray augmentation strategy that operates on the entire sinogram, improving training diversity and robustness to spatially correlated noise without relying on paired or external datasets.

3) We demonstrate the generalizability of our method to multiple real-world imaging scenarios, including low-dose CT and photon-counting CT, achieving high-fidelity denoising across modalities.

## II. Method

### A. Theory of Conjugate Symmetry

A CT scanner acquires measurements of the same object over a full 360-degree rotation. In fan-beam geometry, let $\beta$ denote the source rotation angle and $\gamma$ the detector angle relative to the central ray. Following the standard coordinate definitions in [45], the projection at source angle $\beta$ and detector angle $\gamma$ is then expressed as

$$p(\beta,\gamma) = \int_{\mathbb{R}^2} f(\boldsymbol{x})\delta\left(\left(\boldsymbol{x} - \boldsymbol{s}(\beta)\right) \cdot \boldsymbol{d}_{\perp}(\beta,\gamma)\right) d\boldsymbol{x}, \qquad (1)$$

where $f(\boldsymbol{x})$ denotes the object attenuation, $\boldsymbol{s}(\beta)$ is the source position, $\boldsymbol{d}_{\perp}(\beta,\gamma)$ is the unit vector perpendicular to the ray direction, and $\delta(\cdot)$ is the Dirac delta function.

Each X-ray $(\beta,\gamma)$ has a conjugate counterpart at $(\beta+\pi+2\gamma,-\gamma)$ that traverses the same path through the object, ideally detecting identical signals,

$$p(\beta,\gamma)=p(\beta+\pi+2\gamma,-\gamma). \quad (2)$$

Figure 1(a) illustrates the concept of conjugate X-rays using a fan-beam geometry for clarity. The schematic rays from two different views are shown in blue and green, while their conjugate pair is highlighted in red. Along this path, X-rays from the two opposing views pass through the same region of the object. These two views correspond to the blue and green lines in the sinogram. The red dot in the sinogram marks the conjugate value pair.

We assume that the measured photon counts follow Poisson distribution

$$I(\beta,\gamma)\sim Poisson\left(I_0 e^{-p(\beta,\gamma)}\right), \quad (3)$$

and the log-transformed projection is

$$\hat{p}(\beta,\gamma)=-ln\frac{I(\beta,\gamma)}{I_0}=p(\beta,\gamma)+\ \epsilon(\beta,\gamma), \quad (4)$$

where $\epsilon(\beta,\gamma)$ represents noise. Differences in the noise distributions of the two measurements cause the conjugate values to differ

$$\hat{p}(\beta,\gamma)\neq\hat{p}(\beta+\pi+2\gamma,-\gamma). \quad (5)$$

However, since each pair of conjugate projections represents the same X-ray path, their expected values should be identical

$$\mathbb{E}[\hat{p}(\beta,\gamma)]=\mathbb{E}[\hat{p}(\beta+\pi+2\gamma,-\gamma)], \quad (6)$$

but differ in realization due to independent noise distributions. To exploit this property, we introduce a conjugate-mapping operator that maps projection values between conjugate pairs

$$\mathcal{M}:\hat{p}(\beta,\gamma)\mapsto\hat{p}(\beta+\pi+2\gamma,-\gamma). \quad (7)$$

Theoretically, mapping a single pair of conjugate projection values can generate a sinogram with a different noise distribution. Given an $M\times N$ sinogram, allowing for $2^{(M\times N)}$ possible sinograms with distinct noise distributions. However, in practice, mapping a single pair has minimal impact, making it unsuitable for training a denoising network. To induce a significant change in the noise distribution, we randomly map a fraction $L$ of all sinogram pixels through their conjugate counterparts, with random seed $k$.

$$\widehat{\boldsymbol{P}}'_{\boldsymbol{L,k}}=\mathcal{M}_{L,k}(\widehat{\boldsymbol{P}}), \quad (8)$$

where $\widehat{\boldsymbol{P}}$ and $\widehat{\boldsymbol{P}}'_{\boldsymbol{L,k}}$ denote the original and augmented sinograms, respectively. This augmentation preserves the underlying structure of the object

$$\mathbb{E}\left[\widehat{\boldsymbol{P}}'_{\boldsymbol{L,k}}\right]=\mathbb{E}\left[\widehat{\boldsymbol{P}}\right]. \quad (9)$$

The original sinogram and the augmented sinograms retain the same structural information but exhibit different noise distributions, enabling effective network training for LDCT denoising. By randomly mapping $L$ pairs of conjugate rays use Eq. (7) and repeating this process $K-1$ times by set different random seed, we generate a dataset of $K$ sinograms: one original and $K-1$ augmented. As illustrated in Fig. 1(b), each of the $K$ sinograms shares the same underlying structural information but exhibits a different noise realization.

It is worth noting that in real practice, strict conjugate rays rarely exist because of the discrete sampling of projection views and the intentional detector shifts used to improve spatial resolution. Therefore, we employ bilinear interpolation to obtain the approximated conjugate values. Since interpolation is widely used in CT reconstruction algorithms, its effect on the overall accuracy is minimal. It is important to distinguish the proposed conjugate-ray mapping from subsampling-based strategy. Subsampling methods construct training pairs by partitioning the image into interleaved subsets, implicitly assuming that neighboring pixels share the same underlying signal. In the sinogram domain, this assumption is violated because adjacent detector channels correspond to different X-ray paths and can exhibit large signal differences. In contrast, conjugate-ray mapping exploits the physical fact that conjugate rays traverse identical paths, guaranteeing that input-target pairs are structurally identical. The left subpanel of Fig. 1(c) shows the absolute residual map between the original sinogram and the augmented sinogram. We can observe that there is no structural difference in the residual image (only noise variations), indicating that interpolation does not introduce noticeable errors. In contrast, the commonly used downsampling strategy produces substantial residual errors, as illustrated in the right subpanel of Fig. 1(c). The center subpanel provides a zoomed-in view that highlights the large discrepancy between the two strategies, further demonstrating the advantage of our proposed augmentation strategy.

*B. Network Training and Inference*

To construct training pairs, as shown in Fig. 1(d), we apply a cyclic shift (roll) operation on the input sinogram sequence: each input sinogram is paired with the next one in the series as its target, and the last sinogram wraps around to the first. For clarity, we denote the original sinogram as $\widehat{\boldsymbol{P}}=\widehat{\boldsymbol{P}}'_{\boldsymbol{L,0}}=\widehat{\boldsymbol{P}}'_{\boldsymbol{L,K}}$. Then the training dataset can be denoted as $\left(\widehat{\boldsymbol{P}}'_{\boldsymbol{L,0}},\widehat{\boldsymbol{P}}'_{\boldsymbol{L,1}},\cdots,\widehat{\boldsymbol{P}}'_{\boldsymbol{L,k}},\cdots,\widehat{\boldsymbol{P}}'_{\boldsymbol{L,K-2}},\widehat{\boldsymbol{P}}'_{\boldsymbol{L,K-1}}\right)$, and the syclic shift dataset can be denoted as $\left(\widehat{\boldsymbol{P}}'_{\boldsymbol{L,1}},\widehat{\boldsymbol{P}}'_{\boldsymbol{L,2}},\ \cdots,\widehat{\boldsymbol{P}}'_{\boldsymbol{L,k+1}},\cdots,\widehat{\boldsymbol{P}}'_{\boldsymbol{L,K-1}},\widehat{\boldsymbol{P}}'_{\boldsymbol{L,K}}\ \right)$. Since all input-target pairs represent the same object with varying noise, the network learns to suppress random noise while preserving structural features. This self-supervised strategy allows effective denoising without requiring clean ground truth. We train $f_\theta$ by minimizing

$$\mathcal{L}(\boldsymbol{\theta})=\frac{1}{K}\sum_{k=0}^{K-1}\left\|f_\theta\left(\widehat{\boldsymbol{P}}'_{\boldsymbol{L,k}}\right)-\widehat{\boldsymbol{P}}'_{\boldsymbol{L,k+1}}\right\|_2^2. \quad (10)$$

During inference, as shown in Fig. 1(e), the same $K$ sinograms in the training dataset are passed through the trained network to produce $K$ denoised outputs. Since all the $K$ sinograms describe the same object, these outputs are then averaged to reduce random errors further, yielding a final denoised sinogram.

The network architecture is designed to be lightweight. As shown in Fig. 1(f), the network applies a 3×3 convolution that expands the input to 48 feature channels followed by a LeakyReLU activation with a negative slope of 0.2. It then stacks two identical residual blocks. Each residual block performs, in order of a 3×3 convolution with 48 filters, LeakyReLU, and a second 3×3 convolution with 48 filters. The block output is added to its input to form a local skip connection. After the two residual blocks, a final 1×1 convolution maps the features back to the original number of channels. All 3×3 convolutions are configured with reflect padding to preserve spatial dimensions and reduce edge artifacts.

For each sinogram, a separate network was trained independently under an unsupervised self-prediction scheme. The input sinogram was normalized to the range from 0 to 1 using its maximum intensity. The network was optimized with the Adam optimizer at a learning rate of $1 \times 10^{-3}$ for 2,000 epochs, and a StepLR scheduler was applied to halve the learning rate every 1,000 epochs. The parameter $K$ was set to 4, which was also used as the batch size to balance reconstruction performance and computational efficiency. The parameter L is set as a fixed percentage of the total sinogram pixels (90% in our experiments), and the random selection of conjugate pairs is performed independently for each augmented sinogram using a different random seed. Here $L$ denotes the pixel-level mapping ratio, not the fraction of distinct conjugate ray pairs; because each mapped pixel is paired with its conjugate, the two definitions differ only by a factor of two in bookkeeping. This ensures that each of the K generated sinograms has a distinct noise realization while preserving the underlying structure. All experiments were implemented in PyTorch.

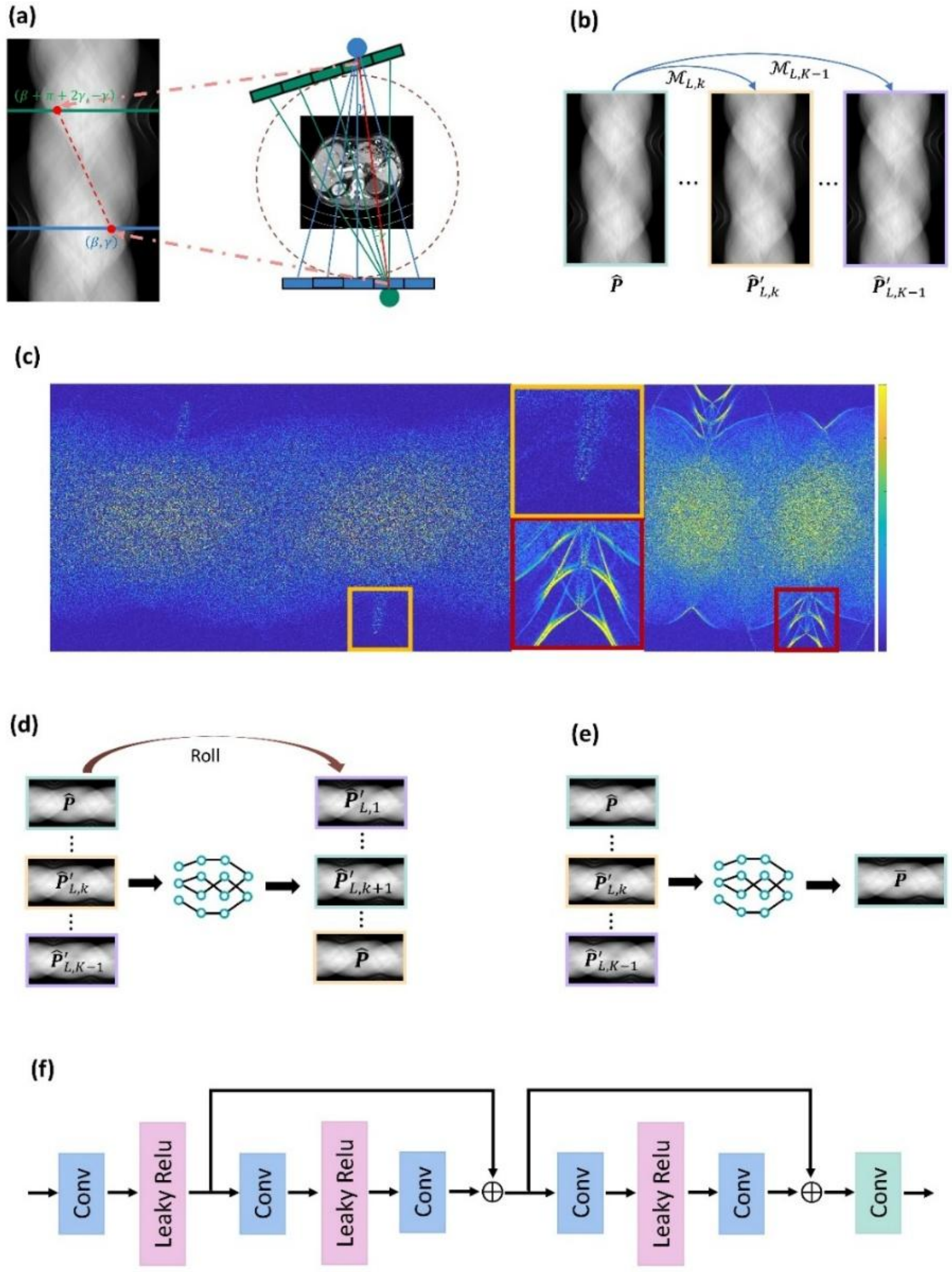

Fig. 1. An overview of our proposed PZL framework. (a) Illustration of conjugate symmetry in CT imaging, where conjugate values share the same expectation but differ due to independent noise. (b) Generation of $K$ sinograms by randomly mapping $L$ pairs of conjugate rays, repeating $K-1$ times with different random seeds. (c) Residual between two generated sinograms; left subpanel is from our conjugate mapping strategy, and right subpanel is from the down-sampling strategy. The center subpanel shows a zoomed view to highlight the large difference. (d) Training via cyclic data rolling. (e) Inference by ensemble averaging of the outputs. (f) The structure of the proposed lightweight convolutional network.

### C. Data Preparation

To enable quantitative evaluation, we first conduct a simulation study. A total of 27 representative slices containing liver lesions from 10 patients are selected from the widely used Mayo Clinic dataset to assess the effectiveness of the proposed method in the liver region [18]. These images are forward-projected into sinograms using a fan-beam geometry. The simulated CT system adopts fan-beam geometry with 1,160 projection views uniformly distributed over 360 degrees. The detector consists of 672 elements. The source-to-isocenter distance is 538.52 mm, and the reconstructed image has 512 × 512 pixels with 0.6 mm × 0.6 mm per pixel. we use Eq. (3) to add Poisson noise in the simulated sinogram and set $I_0$ to $1 \times 10^5$ to simulate the low dose sinogram. This configuration closely mimics a clinical abdominal CT scanner and allows for realistic modeling of projection data and noise.

To evaluate the robustness of our proposed method in real clinical scenarios, we also conduct a helical CT reconstruction study using clinical patient data. The subjects recruited to this study were scheduled for CT-guided lung nodule needle biopsy at Stony Brook University Hospital [46]. The study protocol was approved by the Institutional Review Board, and subjects provided written informed consent. During biopsy, a normal-dose CT scan was performed first, followed by LDCT scans until the needle reached the lung nodule. For both normal and low dose scans, the X-ray tube voltage was set to 120 kVp; the tube current was 100 mAs for the normal-dose scan, and 20 mAs for subsequent low-dose scan. All patients were scanned using a Siemens SOMATOM CT scanner. Raw data were system-calibrated and exported as line integrals, and the helical projections were subsequently rebined into fan-beam geometry. The line integrals include 672 detector elements with width 1.4 mm per element, and 1,160 projection views over a 360 range. The distance from the source to the system center was 570 mm, and the distance from the source to the detector was 1,040 mm. The reconstructed image has 512 × 512 pixels with 0.74 mm × 0.74 mm per pixel.

Beyond conventional CT, we also evaluate our method on a preclinical photon-counting CT dataset. An iohexol solution with 10 mg iodine per milliliter was injected into a mouse, which was then imaged under anesthesia on a photon-counting detector based micro-CT system at the Institute of High Energy Physics, Chinese Academy of Sciences [47]. The system used a 90 kVp X ray source with a 4 mm aluminum filter and a photon counting linear detector with 2,055 pixels at 0.1 mm pitch. The distance from the source to the isocenter was 200.6 mm, and the distance from the source to the detector was 362.8 mm. Each rotation acquired 2,160 projections with a tube current time product of 64.8 mAs. Six energy windows were recorded with thresholds at 28, 34, 40, 47, 56, and 67 keV, producing energy channels of 28 to 34 keV, 34 to 40 keV, 40 to 47 keV, 47 to 56 keV, 56 to 67 keV, and 67 to 90 keV. Given the six energy channels and a peak voltage of 90 kVp, the dose per channel is inherently low. The cone beam acquisition covered 60 slices along the longitudinal direction; to reduce noise we extracted the central four slices and merged them into a single slice for analysis. The reconstructed images form 512×512×6 volumes within plane resolution of 210×210 $um^2$.

### D. Comparison Methods and Evaluation Metrics

We compare denoising against four state-of-the-art denoisers: Noise2Fast, ZS-Noise2Noise, Noise2Noise and Noise2Clean. For fairness, we strictly adhere to the official implementations or published configurations whenever available.

**Noise2Fast [32].** Noise2Fast accelerates self-supervised denoising by applying a checkerboard downsampling strategy to generate four spatially interlaced sub-images from a single noisy input. These sub-images act as independent training samples for rapid convergence, while the original noisy image is retained for monitoring and consistency evaluation.

**ZS-Noise2Noise [33].** ZS-Noise2Noise builds upon Noise2Fast by introducing a self-consistency loss that stabilizes training and eliminates the need for early stopping. The method adapts zero-shot learning to denoising by optimizing the network directly on each test image without pre-training or external data.

**Noise2Noise.** Noise2Noise represents a supervised training paradigm in which paired noisy images with the same noise statistics are used as input and target. In this study, simulated noisy pairs were generated to mimic low-dose conditions. The network structure and training setup were kept consistent with our proposed method to ensure a fair comparison.

**Noise2Clean.** Noise2Clean follows the conventional supervised denoising framework, where a clean image serves as the target for each noisy input. The network structure and training setup were also kept consistent with our proposed method to ensure a fair comparison. This method serves as a supervised reference baseline trained using clean target sinograms.

For a fair comparison, Noise2Fast and ZS-Noise2Noise were evaluated using their official implementations and recommended hyperparameters, without using the PZL backbone, as their training schemes are coupled to their respective architectures. Noise2Noise and Noise2Clean used the same lightweight network architecture and training schedule as PZL to ensure a fair comparison. To evaluate the effectiveness of the proposed approach, we perform a comprehensive evaluation that includes both qualitative visualization and quantitative analysis. For quantitative assessment, we use two widely adopted image quality metrics: peak signal-to-noise ratio (PSNR) and structural similarity index measure (SSIM).

## III. Results

### A. Simulation Study in Liver Lesion Dataset

To illustrate the denoising performance of the selected networks, three representative slices are shown in Fig. 2. The zoomed regions of interest (ROIs), marked by red rectangles, are displayed at the lower left of each image, while the corresponding absolute residual maps are provided at the lower right. The LDCT images are dominated by stochastic noise, which obscures fine anatomical structures. Noise2Fast produces smoother results and reduces noise but tends to over-smooth anatomical details, as evident in the residual maps, and introduces secondary artifacts, especially in the top-row ROI. ZS-Noise2Noise method with consistency loss better preserves structural information and effectively suppresses noise, showing a clear improvement over LDCT. However, residual noise still hinders the visibility of low-contrast details, particularly in the middle-row images. Noise2Noise approach also reduces noise to some extent, but residual artifacts remain visible. While Noise2Noise may theoretically match Noise2Clean's performance with infinite data, its effectiveness is limited when training data are scarce—as in this single-image case. The supervised Noise2Clean model trained with normal-dose CT references achieves high-quality reconstructions with clearly distinguishable lesions. Notably, our proposed PZL method attains comparable visual quality to Noise2Clean, despite requiring no reference data for training.

Quantitative results summarized in Table I further confirm these observations. As expected, Noise2Fast achieves higher scores than ZS-Noise2Noise and Noise2Noise but exhibits over-smoothing. The supervised Noise2Clean model provides substantial improvements, particularly in SSIM. Among the zero-shot methods, the proposed PZL achieves the best performance on both metrics. Its PSNR and SSIM are comparable to those of Noise2Clean (PSNR 33.93±1.65 dB vs. 33.92±1.80 dB; SSIM 83.84±2.70% vs. 83.17±2.71%), while requiring no reference data for training. This improvement arises from the conjugate-ray mapping strategy, which enforces physically consistent constraints within the sinogram domain.

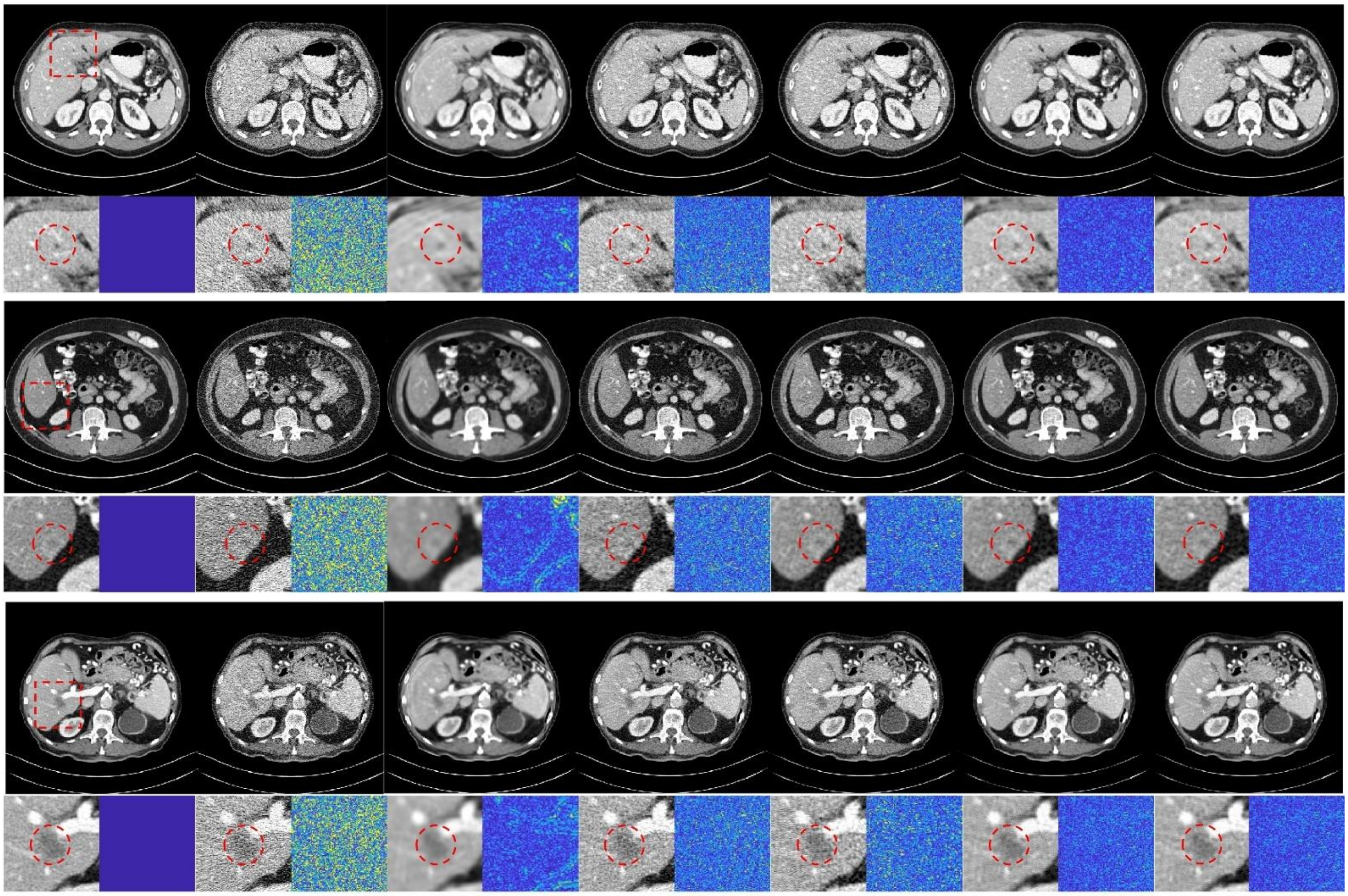

Fig. 2. Three representative transverse abdominal CT images to demonstrate low-attenuation liver lesions (highlighted by red rectangles). From left to right, the columns are the normal-dose CT (reference), LDCT, and the results of Noise2Fast, ZS-Noise2Noise, Noise2Noise, Noise2Clean, and the proposed PZL method. The zoomed ROIs from the red rectangles are shown at the lower left of each image, and the corresponding absolute residual maps are displayed at the lower right. The liver lesions within the ROIs are indicated by red circles. The display window is set to [-125, 225] HU for all reconstructed images and [0, 100] HU (blue-to-yellow scale) for the absolute residual maps.

Table I. PSNR and SSIM on different methods.

| Methods | LDCT | Noise2Fast | ZS-N2N | Noise2Noise | Noise2Clean | PZL |
|---|---|---|---|---|---|---|
| PSNR | 27.46±2.04 | 33.20±2.12 | 32.95±1.59 | 32.15±1.99 | 33.92±1.80 | **33.93**±1.65 |
| SSIM(%) | 68.11±5.52 | 81.00±4.56 | 79.03±3.16 | 76.41±4.61 | 83.17±2.71 | **83.84**±2.70 |

### *B. Parameter Sensitivity and Performance Analysis*

To examine the robustness and behavior of the proposed PZL framework, we conducted a detailed sensitivity analysis on two key parameters introduced in Section II: $L$ to control the proportion of conjugate-ray mappings, and $K$ to determine the number of generated sinograms per input. Both quantitative metrics and qualitative reconstructions are summarized in Fig. 3.

**Effect of $L$.** Parameter $L$ governs the fraction of conjugate-ray pairs reassigned during training, effectively determining the degree of pixel-level mapping between opposing projection bins. To evaluate its impact, we varied $L$ from 10 % to 90 % of the total sinogram pixels, generating target sinograms by randomly reassigning the corresponding conjugate values. As shown in Fig. 3(a), even a small mapping ratio ($L$ = 10 %) enables stable denoising with clear suppression of streak and granular noise while preserving structural details. Increasing $L$ does not produce a conspicuous visual difference. Performance improves monotonically with $L$ and saturates beyond $L$=50%; gains from 50% to 90% are marginal (ΔPSNR < 0.2 dB), and $L$=90% gives the highest mean values. We therefore adopt $L$=90% as default, noting the method is insensitive to $L$ above a moderate threshold.

**Effect of $K$.** Parameter $K$ specifies the number of augmented sinograms generated from each input by applying distinct random conjugate mappings. To assess its influence, $K$ varied from 2 to 8 while keeping other parameters fixed. The corresponding visual results are presented in Fig. 3(b), and the quantitative performance is summarized in Fig. 3(c) (right). With $K = 2$, the network already achieves competitive denoising results, outperforming several existing self-supervised baselines in both PSNR and SSIM. Increasing $K$ to 4 yields a notable improvement in PSNR, suggesting that multiple independent mappings enhance the statistical diversity of training data and stabilize optimization. However, further increasing $K$ beyond 4 yields negligible improvement while extending longer training times, indicating a balance point between diversity and computational efficiency.

Overall, these experiments demonstrate that the proposed framework is robust to a wide range of parameter settings. The combination of $L$=90% and $K = 4$ achieves an optimal trade-off between performance and efficiency and is therefore adopted as the default configuration in subsequent evaluations.

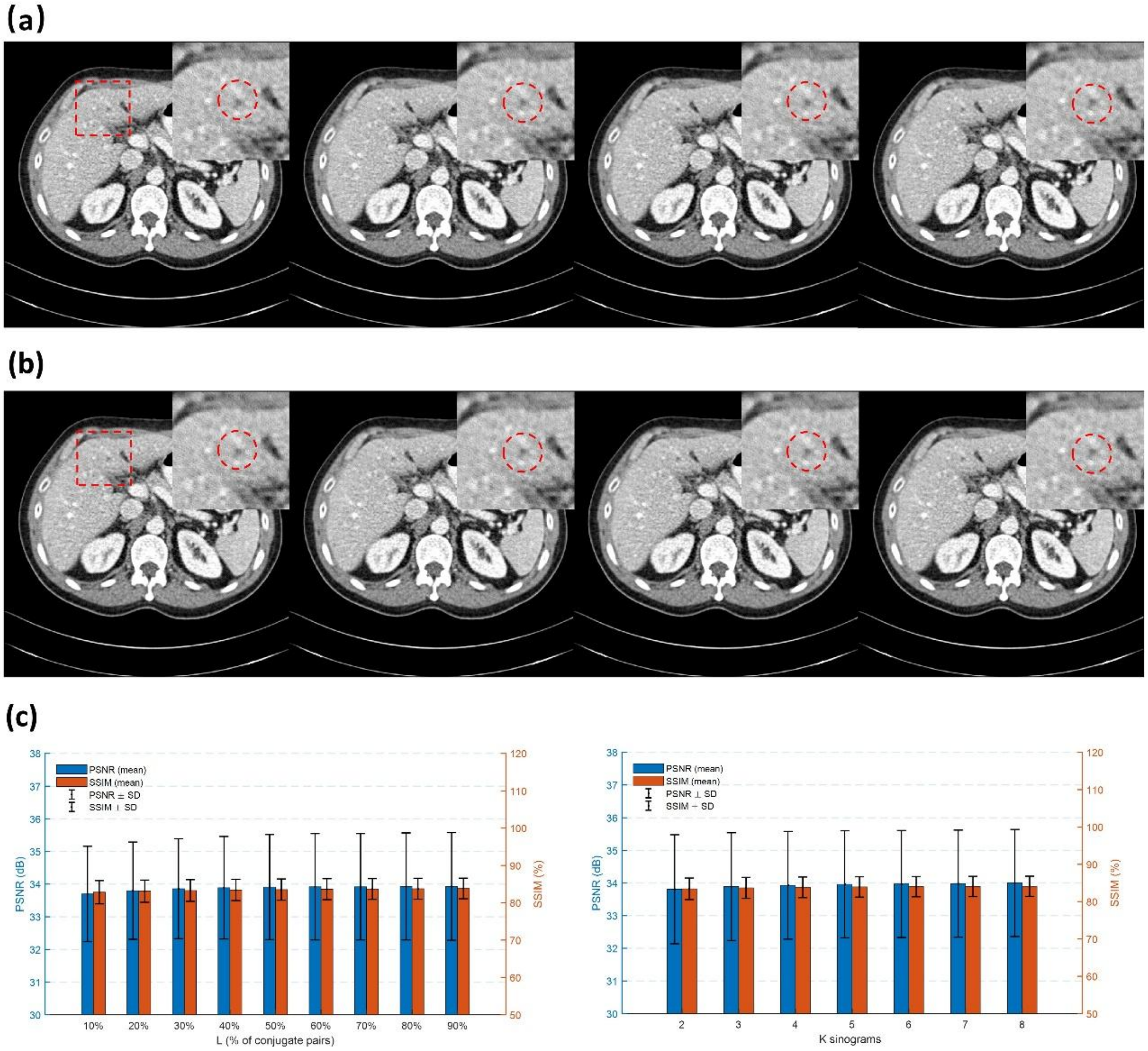

Fig. 3. Results under different parameter settings. (a) From left to right, the images are the reconstructions obtained with parameter $L$ set to 10%, 40%, 70%, and 90% of the total sinogram pixels. (b) From left to right, the images are the results with parameter $K$ set to 2, 4, 6, and 8. The ROIs outlined by red rectangles are enlarged at the top-right corner of each image, and the lesions are marked by red circles. The display window is [-125, 225] HU. (c) Quantitative evaluation of PSNR and SSIM as $L$ varies from 10% to 90% and $K$ from 2 to 8.

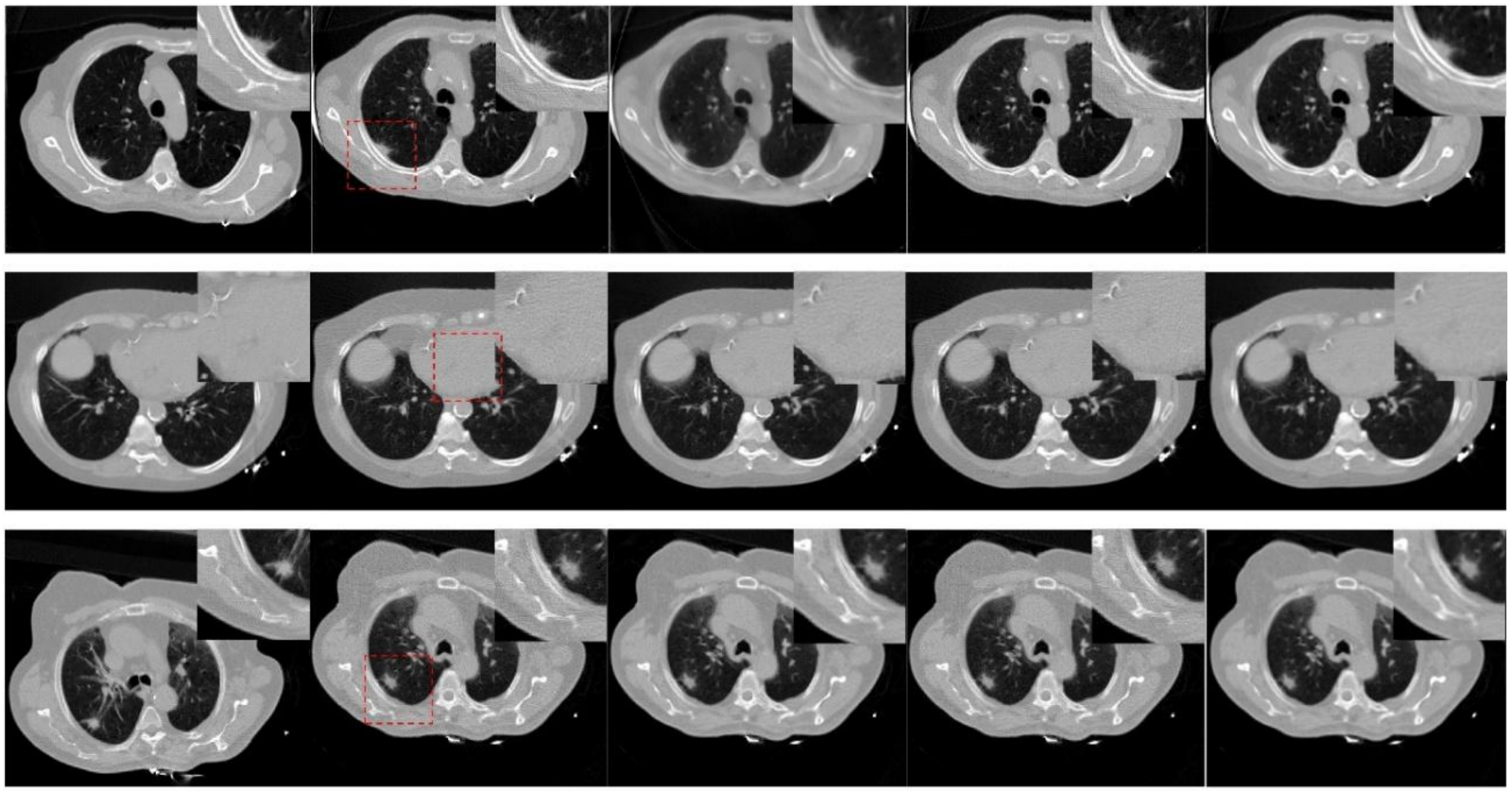

Fig. 4. Visual inspection of denoising performance on clinical helical LDCT scans. From top to bottom the rows are the results from three different patients. From left to right, the columns are the normal-dose CT image (100 mAs) and the LDCT (20 mAs) reconstructions obtained using filtered back projection (FBP), Noise2Fast, ZS-Noise2Noise, and the proposed PZL method, respectively. The display window is [−125, 225] HU. Zoomed ROIs are shown in the upper right corners of each image to highlight texture preservation and noise suppression performance.

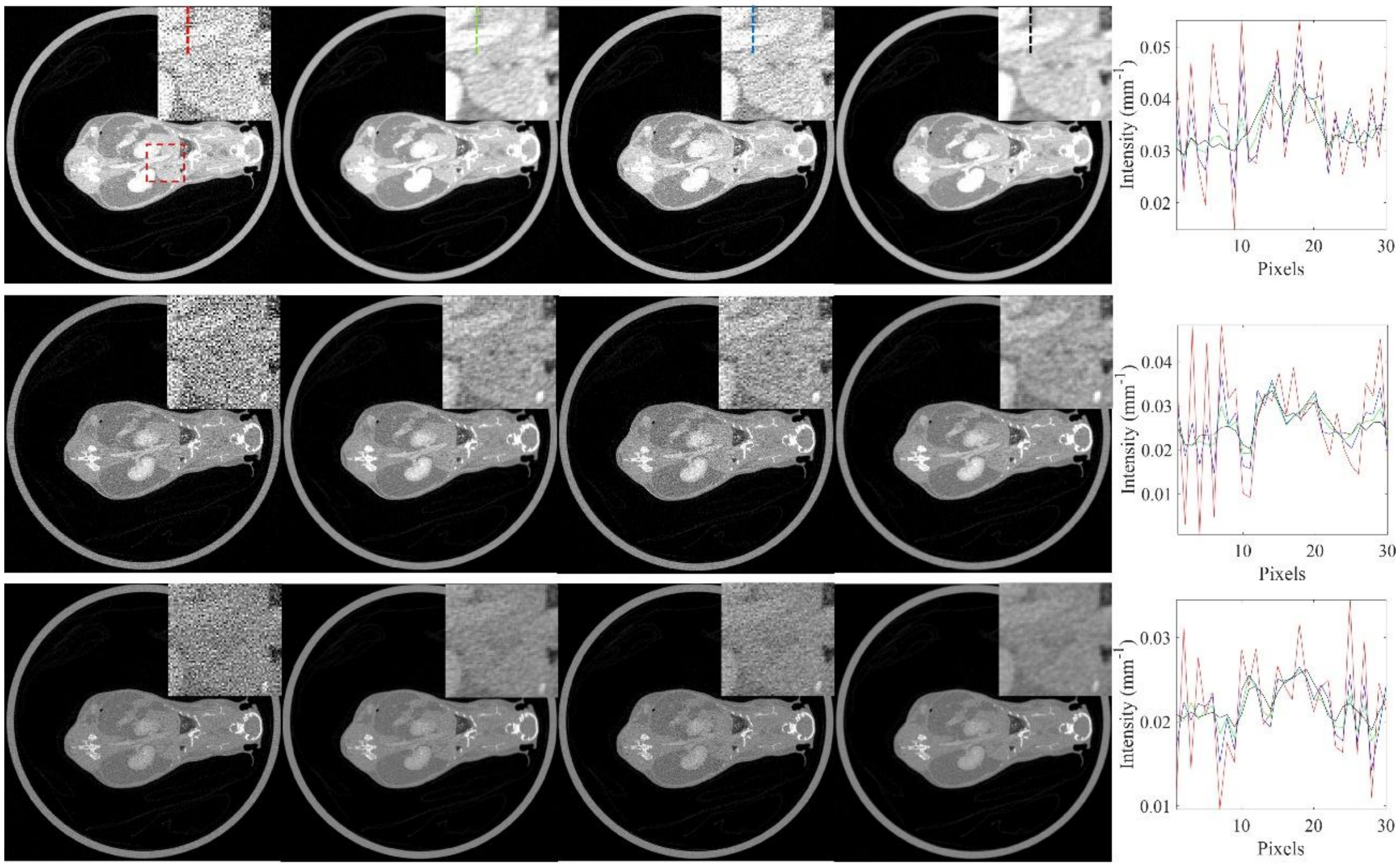

Fig. 5. Qualitative evaluation on preclinical photon-counting CT. From top to bottom the rows are the representative results from three different channels (2, 4, and, 6). From left to right, the columns are reconstructions obtained using FBP, Noise2Fast, ZS-Noise2Noise, and the proposed PZL method, respectively. The display window is set to [0,0.04] $mm^{-1}$. Zoomed ROIs are shown in the upper right corners of each image. The last column is the profiles along the dash line in the zoomed ROIs.

### *C. Real Clinical Low-Dose CT*

To evaluate the clinical applicability of the proposed method, qualitative comparisons were conducted on real helical LDCT scans from three different patients, as shown in Fig. 4. The FBP reconstructions from 20 mAs acquisitions exhibit modest quantum noise and minimal streaking; the lung parenchyma remains largely clean due to the inherently high air–tissue contrast, and soft-tissue details are generally well preserved. This property explains why LDCT is widely adopted for lung imaging. However, noise artifacts can still be clearly observed, particularly in the magnified regions. The Noise2Fast method effectively suppresses noise; however, it causes noticeable oversmoothing, especially in the first patient case (top row), where fine anatomical details are blurred. This behavior illustrates the instability of Noise2Fast when applied to real clinical data, resulting in inconsistent reconstruction quality across different patients. The ZS-Noise2Noise method suppresses noise to a certain extent, but residual noise remains visible, particularly in homogeneous soft-tissue regions, indicating limited denoising capability under real clinical conditions. In contrast, the proposed PZL method achieves the best overall denoising performance, effectively removing streak and granular noise while preserving clear anatomical structures and natural tissue appearance. The reconstructed results remain consistent across different patients, demonstrating the stability and reliability of the proposed framework in clinical LDCT imaging. These results suggest the potential of PZL to improve image quality in low-dose clinical CT.

### *D. Photon-counting CT*

To further assess the robustness and generalizability of the proposed framework, we evaluated its performance on photon-counting CT data, as shown in Fig. 5. The FBP reconstructions exhibit strong quantum noise due to the limited photon counts in each energy channel. The Noise2Fast method effectively suppresses noise. However, it produces mild oversmoothing and residual low-frequency fluctuations, particularly visible in the zoomed-in areas. The ZS-Noise2Noise method suppresses noise to some extent, but residual noise remains apparent, indicating limited robustness under real photon-counting CT conditions. In contrast, the proposed PZL method achieves the cleanest and most visually stable reconstructions across all spectral channels, providing superior noise suppression while preserving fine anatomical structures and soft-tissue continuity. The line-profile plots in the last column further confirm these observations. The intensity curves corresponding to the PZL results are smoother and more consistent than those of competing methods, reflecting improved noise uniformity and structural fidelity. These findings demonstrate that the proposed zero-shot sinogram-domain denoising framework maintains stable performance across different spectral channels, effectively addressing the noise heterogeneity inherent in photon-counting CT imaging.

## IV. Discussion

The experimental results highlight the broader significance of embedding physical redundancy into zero-shot learning for CT denoising. By leveraging conjugate-ray consistency inherent to the acquisition geometry, the proposed framework derives informative learning signals directly from a single sinogram, enabling adaptation to scanner-specific noise characteristics and dose levels without cross-patient training. Embedding physics-derived consistency into the learning process enhances structural reliability and yields denoising quality comparable to a supervised Noise2Clean baseline, while remaining fully zero-shot.

Although the framework performs robustly across different dose levels and modalities, several directions remain open for improvement. The lightweight backbone restricts its ability to capture fine anatomical textures under extremely low-dose conditions. Future work may consider more expressive networks, such as transformer-based architectures or deeper CNNs, together with knowledge distillation techniques to maintain computational efficiency in the zero-shot setting [48]. In addition, the current evaluation is based on two-dimensional slices, while real helical CT contains substantial three-dimensional redundancy along axial and helical directions [49]. Leveraging this volumetric redundancy in a fully three-dimensional zero-shot formulation may further enhance structural consistency and extend the framework’s capability. Finally, we plan to investigate the potential applicability of the framework to other modalities with intrinsic acquisition symmetries, including magnetic resonance imaging (MRI) and positron emission tomography (PET), and to extend it through comprehensive three-dimensional downstream evaluations for tasks such as nodule detection and diagnosis [50].

## V. Conclusion

This study introduces a PZL framework that enables high-fidelity reconstruction in low-dose CT without the need for external training data or clean references. By leveraging the intrinsic conjugate-ray symmetry of CT sinograms, PZL learns directly from a single acquisition to suppress stochastic noise while preserving fine anatomical details. The approach integrates physical consistency with data-efficient learning, offering a practical route toward robust denoising across diverse acquisition protocols and scanner platforms. Comprehensive evaluations on simulated, clinical, and photon-counting CT datasets demonstrate that PZL achieves image quality comparable to the supervised Noise2Clean baseline and the best performance among the evaluated zero-shot methods, highlighting the value of embedding physical priors within self-supervised frameworks. Beyond CT denoising, this physics-aware paradigm provides a general foundation for learning directly from raw measurements in data-limited imaging scenarios, paving the way toward adaptive, self-learning reconstruction systems for next-generation medical imaging. From a clinical translation perspective, the zero-shot nature of PZL offers a practical advantage: it can adapt to individual scans without requiring site-specific training data or retraining a fixed model. However, regulatory considerations remain: as a sinogram-domain processing method, PZL would likely be regulated as a medical device software. Future work should include prospective clinical validation and engagement with regulatory bodies to establish appropriate evaluation standards for zero-shot denoising methods.

## References

[1] D. Ardila, A. P. Kiraly, S. Bharadwaj, B. Choi, J. J. Reicher, L. Peng, et al., “End-to-end lung cancer screening with three-dimensional deep learning on low-dose chest computed tomography,” *Nature medicine*, vol 25, no. 6, pp.954-961, 2019.

[2] C. Wang, J. Shao, Y. He, J. Wu, X. Liu, L. Yang, et al., “Data-driven risk stratification and precision management of pulmonary nodules detected on chest computed tomography,” *Nature Medicine*, vol. 30, no. 11, pp.3184-3195, 2024.

[3] L. R. Koetzier, D. Mastrodicasa, T. P. Szczykutowicz, N. R. van der Werf, A. S. Wang, V. Sandfort, et al., “Deep learning image reconstruction for CT: technical principles and clinical prospects,” *Radiology*, *306*(3), vol. 306, no. 3, p.e221257, 2023.

[4] J. Wang, T. Li, H. Lu, and Z. Liang, “Penalized weighted least-squares approach to sinogram noise reduction and image reconstruction for low-dose X-ray computed tomography,” *IEEE transactions on medical imaging*, vol. 25, no. 10, pp.1272-1283, 2006.

[5] J. Xu, and B. M. Tsui, "Electronic noise modeling in statistical iterative reconstruction," *IEEE Transactions on Image Processing*, vol. 18, no. 6, pp.1228-1238, 2009.

[6] Z. Tian, X. Jia, K. Yuan, T. Pan, and S. B. Jiang, "Low-dose CT reconstruction via edge-preserving total variation regularization," *Physics in Medicine & Biology*, vol. 56, no. 18, p.5949, 2011.

[7] Q. Xu, H. Yu, X. Mou, L. Zhang, J. Hsieh, and G. Wang, "Low-dose X-ray CT reconstruction via dictionary learning," IEEE transactions on medical imaging, vol. 31, no. 9, pp.1682-1697, 2012.

[8] H. Chen, Y. Zhang, M. K. Kalra, F. Lin, Y. Chen, P. Liao et al., "Low-dose CT with a residual encoder-decoder convolutional neural network," *IEEE transactions on medical imaging*, vol. 36, no. 12, pp.2524-2535, 2017.

[9] E. Kang, J. Min, and J.C. Ye, "A deep convolutional neural network using directional wavelets for low-dose X-ray CT reconstruction," *Medical physics*, vol. 44, no. 10, pp. e360-e375, 2017.

[10] J. Adler, and O. Öktem, "Learned primal-dual reconstruction," *IEEE transactions on medical imaging*, vol. 37, no. 6, pp.1322-1332, 2018.

[11] Q. Yang, P. Yan, Y. Zhang, H. Yu, Y. Shi, X. Mou et al., "Low-dose CT image denoising using a generative adversarial network with Wasserstein distance and perceptual loss," *IEEE transactions on medical imaging*, vol. 37, no. 6, pp.1348-1357, 2018.

[12] H. Shan, A. Padole, F. Homayounieh, U. Kruger, R. D. Khera, C. Nitiwarangkul et al., "Competitive performance of a modularized deep neural network compared to commercial algorithms for low-dose CT image reconstruction," *Nature Machine Intelligence*, vol. 1, no. 6, pp.269-276, 2019.

[13] J. He, Y. Wang, and J. Ma, "Radon inversion via deep learning," IEEE transactions on medical imaging, vol. 39, no. 6, pp.2076-2087, 2020.

[14] L. Yang, Z. Li, R. Ge, J. Zhao, H. Si, and D. Zhang, "Low-dose CT denoising via sinogram inner-structure transformer," *IEEE transactions on medical imaging*, vol. 42, no. 4, pp.910-921, 2022.

[15] Q. Gao, Z. Li, J. Zhang, Y. Zhang, and H. Shan, "CoreDiff: Contextual error-modulated generalized diffusion model for low-dose CT denoising and generalization," *IEEE Transactions on Medical Imaging*, vol. 43, no. 2, pp.745-759, 2023.

[16] Z. Song, L. Xue, J. Xu, B. Zhang, C. Jin, J. Yang, "Real-world low-dose CT image denoising by patch similarity purification," *IEEE Transactions on Image Processing*, vol. 34, pp. 196-208, 2024.

[17] S. Kim, B. J. Kim, and J. Baek, "Dose-aware denoising diffusion model for low-dose CT," *Physics in Medicine and Biology*, 2025.

[18] AAPM. (2017). Low Dose CT Grand Challenge. [Online]. Available: http://www.aapm.org/GrandChallenge/LowDoseCT/

[19] Y. Li, X. Sun, S. Wang, Y. Qin, J. Pan, and P. Chen, "Generation model meets swin transformer for unsupervised low-dose CT reconstruction," *Machine Learning: Science and Technology*, vol. 5, no. 2, p. 025005, 2024.

[20] B. J. Nelson, P. Kc, A. Badal, L. Jiang, S. C. Masters, and R. Zeng, "Pediatric evaluations for deep learning CT denoising," *Medical Physics*, vol. 51, no. 2, pp.978-990, 2024.

[21] J. Lehtinen, J. Munkberg, J. Hasselgren, S. Laine, T. Karras, M. Aittala et al., "Noise2Noise: Learning image restoration without clean data," *arXiv preprint arXiv:1803.04189*, 2018.

[22] A. F. Calvarons, "Improved Noise2Noise denoising with limited data," In *Proceedings of the IEEE/CVF Conference on Computer Vision and Pattern Recognition*, pp. 796-805, 2021.

[23] D. Wu, K. Kim, and Q. Li, "Low-dose CT reconstruction with Noise2Noise network and testing-time fine-tuning," *Medical Physics*, vol. 48, no. 12, pp.7657-7672, 2021.

[24] A. M. Hasan, M. R. Mohebbian, K. A. Wahid, and P. Babyn, "Hybrid-collaborative Noise2Noise denoiser for low-dose CT images," *IEEE Transactions on Radiation and Plasma Medical Sciences*, vol. 5, no. 2, pp.235-244, 2020.

[25] M. O. Unal, M. Ertas, and I. Yildirim, "Proj2Proj: self-supervised low-dose CT reconstruction," *PeerJ Computer Science*, vol. *10*, p.e1849, 2024.

[26] A. Krull, T. O. Buchholz, and F. Jug, "Noise2void-learning denoising from single noisy images," In Proceedings of the IEEE/CVF conference on computer vision and pattern recognition, pp. 2129-2137, 2019.

[27] J. Batson, and L. Royer, "Noise2self: Blind denoising by self-supervision," In *International conference on machine learning*, pp. 524-533, 2019.

[28] A. A. Hendriksen, D. M. Pelt, and K. J. Batenburg, "Noise2inverse: Self-supervised deep convolutional denoising for tomography," *IEEE Transactions on Computational Imaging*, vol. *6*, pp.1320-1335, 2020.

[29] S. Yun, U. Jeong, T. Kwon, D. Choi, T. Lee, S. J. Ye et al., "Penalty-driven enhanced self-supervised learning (Noise2Void) for CBCT denoising," In *Medical Imaging 2023: Physics of Medical Imaging* (Vol. 12463, pp. 464-469). SPIE, 2023.

[30] T. Huang, S. Li, X. Jia, H. Lu, and J. Liu, "Neighbor2neighbor: Self-supervised denoising from single noisy images," In *Proceedings of the IEEE/CVF conference on computer vision and pattern recognition*, pp. 14781-14790, 2021.

[31] P. Wei, L. Wang, J. Gan, X. Shi, and M. Shang, "Incorporation of Structural Similarity Index and Regularization Term into Neighbor2Neighbor Unsupervised Learning Model for Efficient Ultrasound Image Data Denoising," *Applied Sciences*, vol. 14, no. 7, p.7988, 2024.

[32] J. Lequyer, R. Philip, A. Sharma, W. H. Hsu, and L. Pelletier, "A fast blind zero-shot denoiser," *Nature Machine Intelligence*, vol. 4, no. 11, pp.953-963, 2022.

[33] Y. Mansour, and R. Heckel, "Zero-shot noise2noise: Efficient image denoising without any data," In *Proceedings of the IEEE/CVF Conference on Computer Vision and Pattern Recognition*, pp. 14018-14027, 2023.

[34] X. Yin, Q. Zhao, J. Liu, W. Yang, J. Yang, G. Quan et al., "Domain progressive 3D residual convolution network to improve low-dose CT imaging," *IEEE transactions on medical imaging*, vol. 38, no. 12, pp.2903-2913, 2019.

[35] H. R. Kanan, A. Adelöw, and M. Colarieti-Tosti, "Cross-Domain Reconstruction Network Incorporating Sinogram Sinusoidal-Structure Transformer Denoiser and UNet for Low-Dose/Low-Count Sinograms," *IEEE Transactions on Radiation and Plasma Medical Sciences*, 2025.

[36] H. S. Park, K. Kim, and K. Jeon, "Low-dose CT image reconstruction with a deep learning prior," *IEEE Access*, *vol. 8*, pp.158647-158655, 2020.

[37] X. Liu, Y. Xie, C. Liu, J. Cheng, S. Diao, S. Tan et al., "Diffusion probabilistic priors for zero-shot low-dose CT image denoising," *Medical Physics*, vol. 52, no. 1, pp.329-345, 2025.

[38] A. Parakh, J. Cao, T. T. Pierce, M. A. Blake, C. A. Savage, and A. R. Kambadakone, "Sinogram-based deep learning image reconstruction technique in abdominal CT: image quality considerations," *European radiology*, vol. 31, no. 11, pp.8342-8353, 2021.

[39] L. Melazzini, C. Bortolotto, L. Brizzi, M. Achilli, N. Basla, A. D'Onorio De Meo et al., "AI for image quality and patient safety in CT and MRI," *European Radiology Experimental*, vol. 9, no.1, p.28, 2025.

[40] J. F. Barrett, and N. Keat, "Artifacts in CT: recognition and avoidance," *Radiographics*, vol. 24, no. 6, pp.1679-1691, 2004.

[41] X. Dong, M. Petrongolo, T. Niu, and L. Zhu, "Low-dose and scatter-free cone-beam CT imaging using a stationary beam blocker in a single scan: phantom studies," *Computational and mathematical methods in medicine*, vol. 2013, no. 1, p.637614, 2013.

[42] Y. Zhang, X. Mou, G. Wang, H. Yu, "Tensor-based dictionary learning for spectral CT reconstruction," *IEEE Transactions on Medical Imaging*, vol. 36, no. 1, pp. 142-154, 2016.

[43] Y. Shi, Y. Gao, Y. Zhang, J. Sun, X. Mou, Z. Liang, "Spectral CT reconstruction via low-rank representation and region-specific texture preserving Markov random field regularization," *IEEE Transactions on Medical Imaging*, vol. 39, no. 10, pp. 2996-3007, 2020.

[44] E. Ström, M. Persson, A. Eguizabal, and O. Öktem, "Photon-counting CT reconstruction with a learned forward operator," *IEEE Transactions on Computational Imaging*, *vol. 8*, pp.536-550, 2022.

[45] A. C. Kak, and M. Slaney, "*Principles of computerized tomographic imaging,*" Society for Industrial and Applied Mathematics, 2001.

[46] Y Gao, J. Tan, Y. Shi, H. Zhang, S. Lu, A. Gupta et al., "Machine learned texture prior from full-dose CT database via multi-modality feature selection for Bayesian reconstruction of low-dose CT," *IEEE transactions on medical imaging*, vol. 42, no. 11, pp.3129-3139, 2021.

[47] Y. Shi, Y. Gao, Q. Xu, Y. Li, X. Mou, Z. Liang, "Learned tensor neural network texture prior for photon-counting CT reconstruction," *IEEE Transactions on Medical Imaging*, vol. 43, no. 11, pp. 3830-3842, 2024.

[48] J. Wang, Y. Tang, Z. Wu, Q. Du, L. Yao, X. Yang et al., "A self-supervised guided knowledge distillation framework for unpaired low-dose CT image denoising," *Computerized medical imaging and graphics*, vol. *107*, p.102237, 2023.

[49] X. Mou, and J. Duan, "Exploring the redundancy of Radon transform using a set of partial derivative equations: could we precisely reconstruct the image from a sparse-view projection without any image prior?" *Physics in Medicine & Biology*, vol. 70, no. 11, p.115011, 2025.

[50] Z. Liang, S. Chang, Y. Gao, W. Cao, L. R. Kuo, M. J. Pomeroy et al., "Leveraging prior knowledge in machine intelligence to improve lesion diagnosis for early cancer detection," *Medical Physics*, 2025.